\documentclass[conference]{IEEEtran}
\ifCLASSINFOpdf
\else
\fi
\ifCLASSOPTIONcompsoc
  \usepackage[caption=false,font=normalsize,labelfont=sf,textfont=sf]{subfig}
\else
  \usepackage[caption=false,font=footnotesize]{subfig}
\fi
\usepackage[cmex10]{amsmath}
\ifCLASSINFOpdf
   \usepackage[pdftex]{graphicx}
\else
   \usepackage[dvips]{graphicx}
\fi
\usepackage{amssymb}
\usepackage{epstopdf}
\usepackage{graphicx}
\usepackage[caption=false]{subfig}

\usepackage{enumerate}
\usepackage{slashbox}

\usepackage{verbatim}

\usepackage{microtype}
\DisableLigatures{encoding = *, family = * }

\usepackage[nolist]{acronym}
\begin{acronym}
	\acro{AAA}{Authentication, Authorization and Accounting}
	\acro{ACL}{Access Control List}
	\acro{AES-CCM}{Advanced Encryption Standard in Counter Mode with a Cipher Block Chaining Message Authentication Code}
	\acro{AKI}{Accountable Key Infrastructure}
	\acro{API}{Application Programming Interface}
    \acro{BSM}{Basic Safety Message}
    \acro{BYOD}{Bring Your Own Device}
	\acro{C2C-CC}{Car2Car Communication Consortium}
	\acro{C2C}{Car-to-Car}
	\acro{C2I}{Car-to-Infrastructure}
	\acro{CA}{Certification Authority}
	\acro{CN}{Common Name}
	\acro{CAM}{Cooperative Awareness Message}
	\acro{CIA}{Confidentiality, Integrity and Availability}
	\acro{CRL}{Certificate Revocation List}
	\acro{CDN}{Content Delivery Network}
	\acro{COCA}{Cornell OnLine Certification Authority}
	\acro{CSR}{Certificate Signing Request}
	\acro{DAA}{Direct Anonymous Attestation}
	\acro{DDoS}{Distributed Denial of Service}
	\acro{DDH}{Decisional Diffie-Helman}
	\acro{DENM}{Decentralized Environmental Notification Message}
	\acro{DHT}{Distributed Hash Table}
	\acro{DoS}{Denial of Service}
	\acro{DoT}{Department of Transportation}
	\acro{DPA}{Data Protection Agency}
	\acro{ECU}{Electronic Control Unit}
	\acro{ECA}{Enrollment \acl{CA}}
	\acro{ETSI}{European Telecommunications Standards Institute}
	\acro{ECDSA}{Elliptic Curve Digital Signature Algorithm}
	\acro{ECIES}{Elliptic Curve Integrated Encryption Scheme}
	\acro{DL/ECIES}{Discrete Logarithm and Elliptic Curve Integrated Encryption Scheme}
	\acro{ECC}{Elliptic Curve Cryptography}
	\acro{EVITA}{E-safety Vehicle Intrusion protected Applications}
	\acro{FOT}{Field Operational Testing}
	\acro{FPGA}{Field-Programmable Gate Array}
	\acro{GPA}{Global Passive Adversary}
	\acro{GN}{GeoNetworking}
	\acro{GS}{Group Signatures}
	\acro{GM}{Group Manager}
	\acro{GBA}{Generic Bootstrapping Architecture}
	\acro{GUI}{Graphic User Interface}
	\acro{HCA}{Higher-level \acl{CA}}
	\acro{HSM}{Hardware Security Module}
	\acro{HTTP}{Hypertext Transfer Protocol}
	\acro{IEEE}{Institute of Electrical and Electronics Engineers}
	\acro{IETF}{Internet Engineering Task Force}
	\acro{IoT}{Internet of Things}
	\acro{ITS}{Intelligent Transport Systems}
	\acro{IT}{Information Technologies}
	\acro{IMSI}{International Mobile Subscriber Identity}
	\acro{IMEI}{International Mobile Station Equipment Identity}
	\acro{IdP}{Identity Provider}
	\acro{IDS}{Intrusion Detection System}
	\acro{ISP}{Internet Service Provider}
	\acro{LA}{Linkage Authority}
	\acro{LEA}{Law Enforcement Agency}
	\acro{LTC}{Long Term Certificate}
	\acro{LTCA}{Long Term \acl{CA}}
	\acro{LDAP}{Lightweight Directory Access Protocol}
	\acro{LTE}{Long Term Evolution}
	\acro{LBS}{Location-Based Service}
	\acro{MAC}{Message Authentication Code}
	\acro{MCA}{Message \ac{CA}}
	\acro{MEA}{Misbehavior Evaluation Authority}
	\acro{NoW}{Network on Wheels}
	\acro{OBU}{On-Board Unit}
	\acro{OCSP}{Online Certificate Status Protocol}
	\acro{PCA}{Pseudonym \acl{CA}}
	\acro{POB}{Pseudonym-On-Board}
	\acro{POD}{Pseudonym-On-Demand}
	\acro{PDP}{Policy Decision Point}
	\acro{PEP}{Policy Enforcement Point}
	\acro{PIR}{Private Information Retrieval}
	\acro{PKI}{Public-Key Infrastructure}
	\acro{PRECIOSA}{Privacy Enabled Capability in Co-operative Systems and Safety Applications}
	\acro{PRESERVE}{Preparing Secure Vehicle-to-X Communication Systems}
	\acro{P2P}{peer-to-peer}
	\acro{PS}{Participatory Sensing}
	\acro{RA}{Resolution Authority}
	\acro{REST}{Representational State Transfer}
	\acro{RBAC}{Role Based Access Control}
		\acro{RCA}{Root \ac{CA}}
	\acro{RCA}{Root Certification Authority}
	\acro{RSU}{Roadside Unit}
	\acro{SAML}{Security Assertion Markup Language}
	\acro{SAS}{Sample Aggregation Service}
	\acro{SCMS}{Security Credential Management System}
	\acro{SCORE@F}{Système COopératif Routier Expérimental Français}
	\acro{SDSI}{Simple Distributed Security Infrastructure}
	\acro{SeVeCom}{Secure Vehicle Communication}
	\acro{SIT}{Sichere Informationstechnologie}
	\acro{SoA}{Service-oriented-Approach}
	\acro{SSO}{Single-Sign-On}
	\acro{SSL}{Secure Sockets Layer}
	\acro{SOAP}{Simple Object Access Protocol}
	\acro{TACK}{Temporary Anonymous Certified Key}
	\acro{TS}{Task Service}
	\acro{TLS}{Transport Layer Security}
	\acro{TPM}{Trusted Platform Module}
	\acro{TTP}{Trusted Third Party}
	\acro{TVR}{Ticket Validation Repository}
	\acro{URI}{Uniform Resource Identifier}
	\acro{VANET}{Vehicular Ad-hoc Network}
	\acro{V2I}{Vehicle-to-Infrastructure}
	\acro{V2V}{Vehicle-to-Vehicle}
	\acro{V2X}{\ac{V2V} and/or \ac{V2I}}
	\acro{VC}{Vehicular Communication}
	\acro{VM}{Virtual Machine}
	\acro{VSS}{\ac{VC} Security Subsystem}
	\acro{WAVE}{Wireless Access in Vehicular Environments}
	\acro{WSDL}{Web Services Discovery Language}
	\acro{W3C}{World Wide Web Consortium}
	\acro{V}{Vehicle}
	\acro{VANET}{Vehicular Ad-hoc Network}
	\acro{VPKI}{Vehicular Public Key Infrastructure}
	\acro{WS}{Web Service}
	\acro{WoT}{Web of Trust}
	\acro{WSACA}{\ac{WAVE} Service Advertisement \ac{CA}}
	\acro{XML}{Extensible Markup Language}
	\acro{XACML}{eXtensible Access Control Markup Language}
	\acro{3G}{3rd Generation}
\end{acronym}

\usepackage{bibentry}

\begin{document}

\title{{\LARGE The Key to Intelligent Transportation:} \\Identity and Credential Management in \\ \acl{VC} Systems}


\author{\IEEEauthorblockN{Mohammad Khodaei and Panos Papadimitratos\\}
\IEEEauthorblockA{Networked Systems Security Group \\
KTH Royal Institute of Technology \\
Stockholm, Sweden \\
\emph{\{khodaei, papadim\}}@kth.se}}

\maketitle

\begin{abstract} 
\ac{VC} systems will greatly enhance intelligent transportation systems. But their security and the protection of their users' privacy are a prerequisite for deployment. Efforts in industry and academia brought forth a multitude of diverse proposals. These have now converged to a common view, notably on the design of a security infrastructure, a \ac{VPKI} that shall enable secure conditionally anonymous \ac{VC}. Standardization efforts and industry readiness to adopt this approach hint to its maturity. However, there are several open questions remaining, and it is paramount to have conclusive answers before deployment. In this article, we distill and critically survey the state of the art for identity and credential management in \ac{VC} systems, and we sketch a roadmap for addressing a set of critical remaining security and privacy challenges. 
\end{abstract}



%
\IEEEpeerreviewmaketitle

\section{Introduction}
\label{sec:introduction}

\ac{VC} systems can greatly enhance transportation safety and efficiency. Using VC, vehicles can directly communicate [\ac{V2V}] across one or multiple hops, or they can exchange information with \acp{RSU} [\ac{V2I}]. The \ac{CAM} and \ac{DENM} can disseminate valuable information \cite{ETSI-102-638} on potentially dangerous vehicle movement (e.g., collision avoidance), environmental hazards, traffic conditions, and other location-relevant information or even assist regulating traffic \cite{papadimitratos2009vehicular}. 

While the benefit is clear, such a large-scale deployment enabling high-stake applications cannot materialize unless \ac{VC} systems are secure and do not expose users' privacy. For example, only legitimate \ac{VC} on-board equipment should be part of the system, and any modification or forgery of \ac{V2V} or \ac{V2I} messages should be detected. The frequent beaconing of safety \acp{CAM} should not leak the whereabouts of drivers (or passengers) to anyone that deploys a set of commodity radios. These concerns are well understood \cite{papadimitratos2006securing}, and the results of several significant projects and initiatives led to a set of common tools and approaches.

\ac{V2X} communication is protected with the help of public key cryptography where a set of certification authorities (CAs) provide credentials to legitimate vehicles. The credentials are then anonymized, and they are short-lived, which enhances privacy and maintains non-repudiation. The system maintains a mapping of these short-term identities to a long-term identity of the vehicle. These ideas can be found in the first \ac{VC} security architecture \cite{papadimitratos2008secure}, elaborated by the SeVeCom project as well as in subsequent projects [e.g., CAMP \cite{whyte2013security} and \ac{PRESERVE} (http://www.preserve-project.eu/)] and technical standardization documents, notably the IEEE 1609.2 WG (IEEE P1609.2/D12, \textit{Draft Standard for Wireless Access in Vehicular Environments}, January 2012), \ac{ETSI} (\ac{ETSI} TR-102-731, \textit{Intelligent Transport Systems (ITS) Security; Security Services and Architecture} and \ac{ETSI} TR-102-941, \textit{Intelligent Transport Systems (ITS) Security; Trust and Privacy Management}), and harmonization documents [\ac{C2C-CC} (http://www.car-2-car.org/) \cite{bissmeyer2011generic}]. More important, there is a willingness to proceed fast, in the near future, with rolling out the first instances of \ac{VC} protected accordingly.

This can be seen positively, as a vote of confidence, to these available solutions, and there are already \ac{FOT} efforts, seeking to bring them closer to deployment. Furthermore, such security and privacy protection is essentially a baseline. It addresses significant yet specific \ac{VC} problems, but it leaves a range of possible optimizations of secure \ac{VC} protocols, as well as the protection of the in-car network and software and information the whole system relies on (e.g., reliable time and location, and, thus, secure global positioning). The question this raises is: \emph{do we indeed have a cornerstone to build upon secure and privacy-protecting \ac{VC} systems?} More precisely, \emph{do we have all the answers needed to deploy an identity and credential management infrastructure for \ac{VC}}?

To address this question, we critically survey the literature, distilling the latest understanding in academia and industry. A set of open questions remains, and they need to be addressed before deployment. For example, user privacy is not thoroughly protected against infrastructure entities (servers) that are honest but curious, or \ac{VPKI} entities do not enforce policies or are not equipped to preclude special types of misbehavior (disruption or privacy breach related). These issues are primarily technical ones, calling for necessary research. However, they also relate to nontechnical considerations, which affect the systems that will eventually be deployed. In this article, we focus on the technical considerations and briefly discuss how they relate to other factors and the potential deployment scenarios. 

In the rest of this paper, we first provide a brief overview of security and privacy-protecting \ac{VC} systems, focusing mostly on the security infrastructure entities (Sec. \ref{sec:shaping-the-vc-security-infrastructure}). Then, we discuss a number of important \ac{VPKI} components and operation requirements (Sec. \ref{sec:vc-security-infrastructure-development}), critically comparing how the related literature approaches and addresses (or not) these problems. For example, we are concerned with the overall robustness of the secure \ac{VC}, the protection of user privacy, and the practicality of alternative proposals. We find that there are not only distinct approaches but in some cases conflicting views. More important, we realize that there are a number of technical considerations and questions that still have no conclusive answers. We outline those focusing on the deployment of an identity and credential management infrastructure (Sec. \ref{sec:challenges}).

\section{Shaping the \ac{VC} Security Infrastructure}
\label{sec:shaping-the-vc-security-infrastructure}

\begin{table*}[ht] 
	\caption{Cryptographic primitives considered for \ac{VC} Systems.}
    \centering
    \resizebox{18cm}{!} {
    	\begin{tabular}{|l||*{3}{c|}}\hline
    		\backslashbox{\ac{VC} Standards and Harmonization}{Cryptographic Primitives}
    		& \makebox[3em]{\centering {\small \shortstack{Asymmetric Key}}}
    		& \makebox[3em]{\centering \shortstack{Symmetric \\ Key}}
    		& \makebox[3em]{\centering \shortstack{Hash \\ Functions}} \\\hline\hline
    			IEEE 1609.2 & \acs{ECC}: \acs{ECDSA} (P-224 or P-256 curves) or \acs{ECIES} (P-256 curve) & \acs{AES-CCM} & SHA-256 \\ \hline
    			\ac{ETSI} & \acs{ECC}: \acs{ECDSA} (only P-256 curve) or \acs{ECIES} (P-256 curve) & \acs{AES-CCM} & SHA-256 \\ \hline
    			\ac{C2C-CC} & \acs{LTC} (\acs{ECDSA}-256) and Pseudonyms (\acs{ECDSA}-224) & \textemdash & SHA-256 \\ \hline
    	\end{tabular}
    }
    \vspace{2mm}
    \label{table:standard-cryptographic-operations}
\end{table*}

\begin{figure}[t!] 
    \centering
	\includegraphics[width=0.5\textwidth,height=0.5\textheight,keepaspectratio] {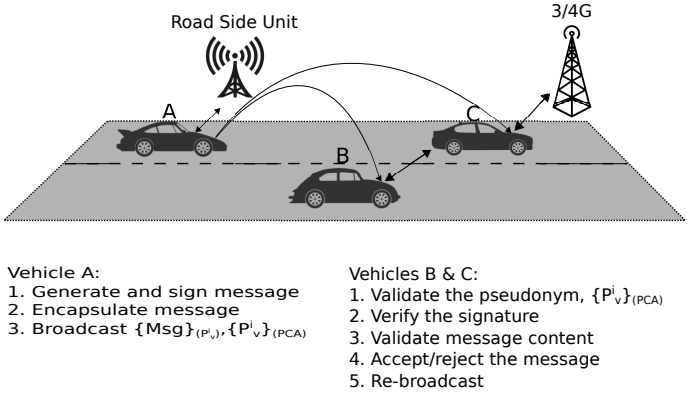}
	\caption{Secure and privacy-protecting \ac{V2X} communication.}
	\label{fig:secure-v2v-v2i-communications}
\end{figure}

Each vehicle is equipped with a set of short-term certificates, termed \emph{pseudonyms}, each with a corresponding short-term private key to sign outgoing messages. Fig. \ref{fig:secure-v2v-v2i-communications} illustrates this: Vehicle $A$ digitally signs outgoing messages (time- and geo-stamped) with the private key, $k^i_v$, corresponding to the pseudonym $P^i_v$ ($\{P^i_v\}_{{(PCA)}}$ represents the pseudonym signed by the pseudonym issuer) and is attached to messages to facilitate verification on the receiver side. Receiving vehicles $B$ and $C$ verify the pseudonym $\{P^i_v\}_{{(PCA)}}$ and validate the signature (assuming they trust the pseudonym issuer discussed below). This process ensures the authenticity and integrity of the message and enables further validation based on its content. At the same time, transmissions by vehicle $A$ do not reveal its identity (as the short-term certificates are anonymised), and messages signed under different pseudonyms (with different private keys) are, in principle, unlinkable. Vehicles switch from one pseudonym to another (not previously used) to achieve unlinkability.

\vspace{1em}
\textbf{Security Infrastructure Entities:} 
A \ac{VPKI} comprises a set of authorities with distinct roles: the \acf{RCA}, the \acf{LTCA}, the \acf{PCA}, and the \acf{RA}. Different proposals may refer to these entities with various names, e.g., CAMP \cite{whyte2013security} refers to the \ac{LTCA} as the \ac{ECA}. The \acp{RCA} are the highest-level authorities certifying \acp{LTCA}, \acp{PCA}, and \acp{RA}. An \ac{LTCA} is responsible for registering vehicles and issuing \acp{LTC}. A \ac{PCA} issues sets of pseudonyms for the registered vehicles. An \ac{RA} can initiate a process to resolve a pseudonym, i.e., identify the long-term identity of the vehicle that used (in a nonrepudiable manner) its short-term keys and credentials.

\begin{figure}[t!] 
    \centering
	\includegraphics[width=0.394\textwidth,height=0.394\textheight,keepaspectratio] {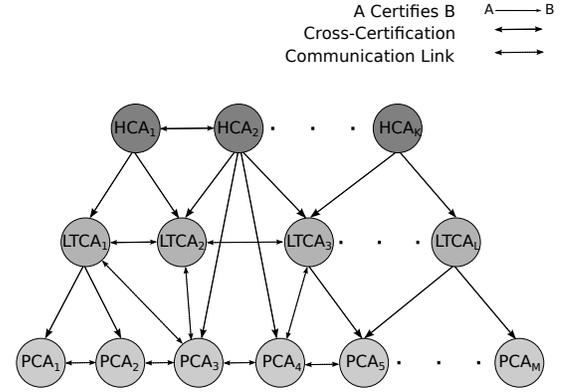} 
	\caption{Hierarchical organization of the \ac{VC} security infrastructure.}
	\label{fig:hierarchy-of-security-authorities}
\end{figure}

\vspace{1em}
\textbf{Trust Associations:} 
Fig. \ref{fig:hierarchy-of-security-authorities} illustrates a generalized hierarchical organization of the \ac{VC} security infrastructure, with multiple \ac{LTCA} and \ac{PCA} entities shown (as those are primarily involved in interactions with the vehicles). 

\ac{VC} systems will be deployed widely, thus one can envision that higher-level CAs (HCAs) could be established to facilitate a trust establishment across distinct parts of the hierarchy. Without loss of generality, let the corresponding numbers of \acsp{HCA}, \acp{LTCA} and \acp{PCA}, be denoted by $K$, $L$ and $M$, so that $K \leq L \leq M$. It is also possible to have direct cross-certification between CAs. There may be direct communication needed among \acsp{CA}, e.g., for lookup operations while issuing credentials. 

\vspace{1em}
\textbf{\ac{VPKI} Structure:}
In \ac{VC} systems, a domain was first described \cite{papadimitratos2006securing} as a set of mobile nodes registered with an authority, with communication independent of administrative or geographical boundaries. Alternatively, a domain could be defined as a (fine- or coarse-grained) geographic region, each with the corresponding \acsp{CA}. The former definition is more general and it is assumed here. A set of vehicles registered with only one \ac{LTCA} can obtain credentials from several \acp{PCA}, subject to compatible policies, as long as the two \acsp{CA} have a trust association. In a multidomain environment, we must determine how to identify available \acp{PCA} and inter-CA trust associations when roaming in a foreign region or interacting with vehicles and roadside infrastructure from a foreign domain; a lightweight directory access protocol service can facilitate this \cite{khodaei2014ScalableRobustVPKI}.

\vspace{1em}
\textbf{Cryptographic Primitives:} 
Table \ref{table:standard-cryptographic-operations} shows cryptographic primitives considered in standardization documents (IEEE P1609.2/D12, \textit{Draft Standard for Wireless Access in Vehicular Environments}, January 2012, ETSI TR-102-731, \textit{Intelligent Transport Systems Security; Security Services and Architecture}, and ETSI TR-102-941, \textit{Intelligent Transport Systems Security; Trust and Privacy Management}), and harmonization efforts. The motivation for \ac{ECDSA} is that it produces shorter signatures than the ones by the Rivest-Shamir-Adleman (RSA) cryptosystem. IEEE 1609.2 considers a \ac{DL/ECIES} to protect communication during the pseudonym acquisition phase, as do other proposals. Symmetric cryptography, i.e. an \ac{AES-CCM}, is proposed for other wireless networking standards IEEE 802.11 or Zigbee/IEEE 802.15.4 (IEEE P1609.2/D12 document).

\section{\ac{VC} Security Infrastructure Development}
\label{sec:vc-security-infrastructure-development}

In most of the literature~\cite{schaub2010v, puca2014, studer2009tacking, vespa2013, bibmeyer2013copra}, including standardization documents (IEEE P1609.2/D12, \ac{ETSI} TR-102-731, and \ac{ETSI} TR-102-941), the \ac{VPKI} entities are assumed fully trustworthy. This is a reasonable assumption; however, recent experience from mobile computing applications and \acp{LBS} shows that applications and services aggressively collect user information. While they may remain trustworthy, not deviating from their protocol specifications and offering reliable services to their users, service providers can be tempted to infer sensitive user information and profile users (e.g., attempting to monetize this by offering customized services), based solely on the prescribed functionality. 

This type of deviation relates to the \emph{honest-but-curious} adversarial model, considered by~\cite{whyte2013security, khodaei2014ScalableRobustVPKI}. In the \ac{VC} context, \ac{VPKI} servers (e.g., \acp{LTCA} and \acp{PCA}) are considered to be honest, complying with security policies and correctly executing protocols, but also curious, seeking to infer user-sensitive information. This can be especially tempting, because a transcript of \ac{V2X} communication (e.g., as it could be collected by a mesh network of \ac{VC}-compatible radios) could be converted into a rich set of user trajectories and profiles if processed with the information that the \ac{VPKI} entities possess. 

This concern, aggravated by a potential spread of the responsibility to run credential and identity providers, is discussed first in this section. In spite of the common understanding that \ac{VPKI} servers should have well-defined distinct roles, safeguarding users from honest-but-curious servers is not trivial and, in most cases, not achieved. At the same time, current \ac{VPKI} designs do not fully prevent abuse of anonymity (or, to be precise, pseudonymity) by malicious (dishonest) clients, i.e., vehicles or \acp{RSU}. This can be seen as a by-product of the role separation. The second part of this section surveys how to improve \acp{VPKI} to render \ac{VC} more trustworthy.

\subsection{Privacy Considerations} \label{subsec:distinct-roles}

Before issuing pseudonyms, the \ac{PCA} either communicates with an \ac{LTCA} to have the requester (vehicle) \ac{VPKI} server  authenticated or authenticates the vehicle itself. Several \ac{VPKI} schemes \cite{whyte2013security, schaub2010v, puca2014, bibmeyer2013copra} follow the former approach, i.e., the \ac{C2C-CC} design proposal where the \ac{PCA} directly communicates with the vehicle's \ac{LTCA}. Another set of \ac{VPKI} schemes \cite{khodaei2014ScalableRobustVPKI, vespa2013, gisdakis2013serosa} proposes an indirect involvement of the \ac{LTCA}, which issues a token to the vehicle that can be presented and verified by the \ac{PCA} before issuing the pseudonyms. 

For both approaches, the motivation is to maintain \emph{distinct roles}, i.e., to separate the long-term identification of the vehicles from their short-term identities (their pseudonyms). A \ac{PCA} should ideally be assured that it serves a legitimate vehicle, without accessing the long-term identity and credentials of the vehicle. On the other hand, the \ac{LTCA} should not know which pseudonyms the vehicle obtained (and for which period). If either of the two happened, then a single \ac{VPKI} entity would breach user privacy: the actions (signed messages) of the vehicle, matched to its pseudonyms, would be linked to each other and the vehicle long-term identity. For the same reason, the overall vehicle-\ac{PCA}-\ac{LTCA} communication should not be accessible by any other observer. 

A common issue for all schemes proposed in the literature is that the \ac{PCA} can trivially link the pseudonyms issued for a vehicle in one pseudonym request. CAMP~\cite{whyte2013security} proposes a proxy-based scheme that the Registration Authority (a proxy to validate, process, and forward pseudonym requests to the \ac{PCA}) shuffles the requests from the vehicles before forwarding them to the \ac{PCA} so that the \ac{PCA} cannot identify which pseudonyms belong to which vehicle. As a result, the registration authority could not link the pseudonyms to the cocoon public keys in the certificate requests, thus to the vehicles, even if it is fed by the transcripts of (pseudo-)anonymized signed messages. 


It is more important to prevent a \ac{PCA} from linking sets of pseudonyms issued for the same vehicle as responses to two or more distinct requests (i.e., pseudonym acquisition protocols). This can be achieved by most proposals as they explicitly preclude, for example, the use of long-term credentials or the use of an available pseudonym for vehicle authentication. 

However, the involvement of the \ac{LTCA} in authenticating the client reveals information. In most proposals, the \ac{LTCA} learns which vehicle requests service from which \ac{PCA}, and, thus, the actual pseudonym acquisition time. This information, along with some default policy data, could make it easy to guess which set of pseudonyms (thus, which set of signed messages) correspond to which vehicle (long-term identity). The problem has been identified only in a recent token-based scheme \cite{khodaei2014ScalableRobustVPKI}. It hides from the \ac{LTCA} the timing information as well as the \ac{PCA} from which the vehicle seeks to obtain its next set of pseudonyms. 

The ramifications of pseudonym lifetime and the use of time information to link pseudonyms are discussed further in the ``challenges'' section. Moreover, decoupling the \ac{LTCA} and the \ac{PCA} knowledge for the sake of privacy raises security and resilience considerations. This is discussed further when we consider revocation (which necessitates maintaining a mapping of short- and long-term identities, unless the scheme is fully anonymized \cite{puca2014}).

\subsection{Resilience considerations}
\label{subsec:sybil-based-misbehavior}

In a multidomain \ac{VC} system with a multiplicity of \acp{PCA}, a compromised vehicle could obtain multiple (sets of) simultaneously valid pseudonyms simply by submitting multiple requests to distinct \acp{PCA}. This presumes a minimal protection to reject spurious requests from the same vehicle and to issue a set of nonoverlapping pseudonyms as a response to each request. With multiple short-term private/public key pairs and the corresponding certificates (pseudonyms), the attacking vehicle could appear as multiple vehicles. It could, for example, inject multiple erroneous hazard notifications and mislead the system (while, perhaps, a single report would not suffice to raise an alarm). 

This ``Sybil-based'' misbehavior, the acquisition of multiple simultaneously valid credentials, is not considered in the \ac{C2C-CC} and CAMP \cite{whyte2013security} designs. A number of other proposals \cite{schaub2010v, puca2014, bibmeyer2013copra, gisdakis2013serosa} do not manage to prevent this without any provision to tie the pseudonym acquisition period to a request. It is not straightforward to have the \ac{LTCA} enforce a policy without revealing information, unless a specific design is put in place to keep information and still allow a policy to be enforced. For example, \cite{khodaei2014ScalableRobustVPKI} does not reveal information to the \ac{LTCA} since the vehicles hide their actual requested interval to obtain pseudonyms with universally fixed lifetimes determined by the \ac{LTCA}. Thus, vehicles can obtain pseudonyms within the requested time interval without revealing the actual pseudonym acquisition period. 

Note that these issues emerge exactly because of the generalization of the system setup and the strengthening of the adversarial model, compared to earlier works, which nonetheless propose alternatives such as the reliance on a \ac{HSM} (ensuring that all signatures are generated under a single valid pseudonym at any time) \cite{papadimitratos2007architecture}.

\subsection{Revocation}
\label{subsec:revocation}

In case of misbehavior, the wrongdoer can be evicted (i.e., prevented from further participating in the system). This is standardized in the Internet, and it is considered for long-term \ac{VC} credentials, the \ac{LTC} of vehicles, and the security infrastructure entities. Nonetheless, what is distinctive here is the multiplicity of short-term credentials used by the vehicles and the need to revoke those as well. Interestingly, the standardization documents (IEEE P1609.2/D12, \ac{ETSI} TR-102-731, and \ac{ETSI} TR-102-941) and harmonization (\ac{C2C-CC}) efforts are inconclusive on that front.

The distinction of \ac{VPKI} roles offers an interesting option: the vehicle can be shunned off by the \ac{LTCA}, which does corroborate its legitimacy to the \ac{PCA} \cite{papadimitratos2007architecture}, thus preventing the vehicle from obtaining any additional pseudonyms. This alone, of course, does not prevent a compromised vehicle from misbehaving while using any pseudonyms it has (and the corresponding private keys) until they expire. The revocation of pseudonyms is necessary to close down this vulnerability window. Consider, for example, the practice outlined in the \ac{C2C-CC} documentation, which recommends preloading the vehicle with approximately 1,500 pseudonyms to be used for one year. An active malicious disruption from an ``insider'' for a significant fraction of a year could be disastrous.

One can reduce this vulnerability by requiring that vehicles interact with the \ac{VPKI} regularly, e.g., once per day or a few times per day, or at least as frequently as the dissemination of revocation information by the \acp{PCA}. Still, within this period, the high-stakes nature of \ac{VC}, possibly risking the well-being of individuals and property, can necessitate a reaction, i.e., revocation of pseudonyms.

The pseudonym revocation can be done by ``traditional'' methods adjusted to the requirements of \ac{VC}. The distribution of \acp{CRL} has been assumed by several proposals \cite{whyte2013security, khodaei2014ScalableRobustVPKI, vespa2013, gisdakis2013serosa}. It was investigated \cite{haas2009design, papadimitratos2008certificate}, along with localized distributed protocols to protect against wrongdoers until they are revoked \cite{moore2008fast}. It was also integrated in recently implemented systems along with a brief comparison with the online certificate status protocol \cite{khodaei2014ScalableRobustVPKI}. The challenge of timely dissemination of credential validity information that does not interfere with vehicle operation remains.

\section{Challenges}
\label{sec:challenges}

Based on and beyond the technical discussion in the previous sections, here we discuss a number of significant challenges for the identity and credential management of fundamental importance toward deploying a secure \ac{VC} system. We extend this discussion by considering a non-technical operational uncertainty at this point.

\vspace{1.1em}
\textbf{Pseudonym Lifetime Policy:} 
The more frequent the changes, the more effective the privacy protection (the higher the unlinkability); ideally, each pseudonym should be used for a single message authentication. However, this could be excessively costly, e.g., if one considers the high-rate safety beaconing (e.g., three to ten beacons per second) and the resultant large numbers of pseudonyms to be provided to each vehicle. Equally important, safety applications necessitate partial linkability, over a period, to facilitate their task. For example, inferring a collision hazard based on logically unlinkable \acp{CAM} would be hard (e.g., needing strictly use of location information) and error prone. Thus, a ``compromise'' was considered early on with partial linkability (over the lifetime of the pseudonym) \cite{papadimitratos2006securing}, while several proposals investigated when/how to change pseudonyms for effective protection. Some current considerations suggest, antidiametrically, using one or a few pseudonyms per day. This divergence of views comes along with the fact that standards and harmonization efforts have not established any guideline for the pseudonym lifetime or other policies. This is clearly a necessity, independently of the flexibility the user would like to enjoy.

A significant consideration that is not pertinent to the privacy-effectiveness tradeoff relates to security. As discussed earlier, without the necessary design, attacking vehicles could amplify the effect of their misbehavior when they obtain multiple simultaneously valid credentials. One approach to prevent this, mentioned previously, is to issue pseudonyms with nonoverlapping lifetimes. This tends to become de facto or implicitly common. However, when combined with flexible access to the \ac{PCA}, as the user needs to, this can undermine unlinkability and timing information can reduce uncertainty and make sets of pseudonyms obtained by the same user linkable (more likely to be). This was recently discovered and a countermeasure was outlined based on enforcing a specific pseudonym lifetime policy \cite{khodaei2014ScalableRobustVPKI}. Again, this emphasizes the need to standardize policies with clear objectives.

\vspace{1.1em}
\textbf{Revocation:} 
As discussed previously, there is no consensus on the need and the method for revocation of pseudonyms. While several \ac{VPKI} proposals address the need of pseudonym revocation \cite{whyte2013security, khodaei2014ScalableRobustVPKI, schaub2010v, vespa2013, gisdakis2013serosa, papadimitratos2008certificate}, standardization bodies and harmonization efforts propose revocation of only long-term credentials, but not the pseudonyms. Moreover, revocation could be necessary for other reasons (e.g., revoking an attribute of a whole class of vehicles \cite{Papadi:C:08}). Essentially, there is a tradeoff between vulnerability and cost, which also rises if one seeks to reduce risk by mandating more frequent vehicle-\ac{VPKI} interactions. This needs to be explored, along with a clear determination of a policy on what events necessitate revocation.

\vspace{1.1em}
\textbf{Extending to Anonymous Authentication Primitives:} 
Although classic public-key cryptography has been a pillar for securing \ac{VC} systems, there have been proposals to leverage anonymous authentication in the context of \ac{VC}. Calandriello et al. \cite{calandriello2007efficient} use group signatures for vehicles to issue on-the-fly pseudonyms with two follow-up investigations in \cite{PapadiCLH-C-08} and \cite{calandriello2011performance}; Studer et al. \cite{studer2009tacking}, Lin et al. \cite{lin2007gsis}, and Lu et al. \cite{lu2008ecpp} also propose the use of group signature protocols with the former using keys as long-term credentials; and F\"{o}rster et al. \cite{puca2014} use zero-knowledge proofs \cite{camenisch2006win} to the \ac{VPKI} infrastructure. A convergence with the standardized approaches could yield significant benefits, and, thus, a recommendation for additional investigations. 

\begin{table}
	\caption{Latency for issuing 100 pseudonyms.}
	\centering
	  \resizebox{0.35\textwidth}{!}
	  {
	    \begin{tabular}{l | *{6}{c} r}
		       & $D_{\ac{PCA}}$ (ms) & $CPU_{\ac{PCA}}$ (GHz)\\ 
		      \hline
		      VeSPA \cite{vespa2013} & 817 & 3.4 \\ 
		      SEROSA \cite{gisdakis2013serosa} & 650 & 2.0 \\ 
		      PUCA \cite{puca2014} & 1,000  & 2.53 \\ 
		      SR-\ac{VPKI} \cite{khodaei2014ScalableRobustVPKI} & 260 & 2.0 \\ 
	    \end{tabular}
	    \label{table:vpki-operations-efficiency-comparison}
      }
\end{table}

\vspace{1.1em}
\textbf{Extensive Experimental Validation:} 
In the light of the \ac{VC} large-scale multidomain environment, the efficiency of the \ac{VPKI} and, more broadly, its scalability are important. Thus far, this has received limited attention, with few schemes evaluating implementation performance. Table \ref{table:vpki-operations-efficiency-comparison} shows the latency to issue 100 pseudonyms in different \ac{VPKI} systems. The dual-core CPU clock is provided only as an indication of the processing power, but clearly a direct comparison is not straightforward with the available information, although the four experimental setups resemble or are at least close to each other. The motivation is to highlight the need for extensive experimental evaluation to ensure the viability (in terms of performance and cost) of the \ac{VPKI} as the \ac{VC} system scales up.

\vspace{1.1em}
\textbf{Operational Challenges:} 
As discussed in the ``Shaping the \ac{VC} Security Infrastructure'' section, a domain is not yet precisely defined across different standardization documents and efforts. The key questions are who will operate the identity and credential provision and how trust relationships will be established. Moreover, policies that determine how to select and certify these \ac{VPKI} entities are necessary. At the same time, \ac{VC} systems will also operate as an extension of the mobile Internet, offering other (e.g, infotainment) services to their users. This would raise the question of how to manage identities and credentials for those applications and how to enable a coexistence of the two ``worlds.''

In conclusion, it is necessary to pave the way for the deployment of secure and privacy-protecting \ac{VC} systems with an identity and credential management infrastructure that builds upon the past multiyear efforts and developed understanding, and addresses a number of open questions to achieve enhanced protection (of the system and its users) and scalability as \ac{VC} becomes ubiquitous. 

\vspace{1em}
\section{Acknowledgement}
\label{sec:acknowledgement}

The research leading to these results received funding from the \acf{PRESERVE} FP7 European project (http://www.preserve-project.eu). 

\vspace{1em}
\textbf{Author Information}

\textbf{Mohammad Khodaei} (khodaei@kth.se) earned his diploma in software engineering from Azad University of Najafabad in Isfahan, Iran, in 2006 and his M.S. degree in information and communication systems security from KTH, Stockholm, Sweden, in 2012. He is currently pursuing his Ph.D. degree at the Networked Systems Security Group, KTH, under the supervision of Prof. Panos Papadimitratos. His research interests include identity and credential managements in vehicular ad hoc networks, the
Internet of Things, and smart cities.

\textbf{Panagiotis (Panos) Papadimitratos} (papadim@kth.se) earned his Ph.D. degree from Cornell University, Ithaca, New York, in 2005. He then held positions at Virginia Tech, \'{E}cole Polytechnique F\'{e}d\'{e}rale de Lausanne, and Politecnico of Torino. He is currently an associate professor at KTH, where he leads the Networked Systems Security Group. His research agenda includes a gamut of security and privacy problems, with emphasis on wireless networks.

\bibliographystyle{IEEEtran}
\bibliography{references}

\end{document}